\documentclass[12pt,preprint]{aastex}
\usepackage{amsmath}

\title{A Simple Model for Understanding the DIM Dust Measurement at Comet 67P/Churyumov-Gerasimenko}

\author{Morris Podolak$^1$\footnote{Corresponding author. e-mail:podolakmorris@gmail.com\ \  fax: 972 36409282}, Alberto Flandes$^{2,4}$, Vincenzo Della Corte$^3$, and Harald Kr\"uger$^4$}
\affil{$^1$Department of Geosciences,\\
Tel-Aviv University, 69978, Israel. \\
$^2$Ciencias Espaciales, Instituto de Geof\'isica, \\
Universidad Nacional Aut\'onoma de M\'exico, Coyoac\'an 04510, M\'exico, D.F. \\
$^3$Istituto di Astrofisica e Planetologia Spaziali, Istituto
Nazionale di Astrofisica (INAF), Via Fosso del Cavaliere, 100,
0133 Rome, Italy. \\
$^4$Max-Planck-Institut f\"ur Sonnensystemforschung, Justus-von-Liebig-Weg 3, 37077 G\"ottingen, Germany. \\
}
\begin{document}

\begin{abstract}
We present a simple model for gas and dust flow from 67P/Churyumov-Gerasimenko that can be used to understand the grain impact observed by the DIM instrument on Philae \citep{kruger15}.  We show how model results when applied to the GIADA measurements \citep{rotundi15, dellacorte2015} can be used, in conjunction with the results found by the MIRO \citep{schloerb2015} and VIRTIS \citep{desanctis2015} instruments to infer surface properties such as surface temperature and surface ice fraction.
\end{abstract}

\section{Introduction}
During the Philae descent to the surface of comet 67P/Churyumov-Gerasimenko, the Dust Impact Monitor (DIM) -aboard the Philae lander- recorded the impact of a cometary grain \citep{kruger15}.  

The DIM instrument is a $\sim 7$\,cm cube with three sides covered with three piezoelectric plates each. DIM is designed to detect the millimeter-sized dust particles that impact on its active surface. Every impact of each dust particle produces a signal from which its maximum voltage amplitude ($U_m$) and the duration of this peak (or contact time, $T_c$) can be related to the size (radius $r$) and the impact speed ($v$) of the particle (assuming given mechanical properties, like its density, $\rho$, and Young's modulus, $Y$) by means of the Hertz impact theory \citep{seidensticker2007,flandes14}. Besides, the onboard DIM electronics is only able to store the $U_m$ and $T_c$ values. 

 The detected particle produced an impact signal with $U_m$=2.45\,mV and $T_c=61\,\mu s$ \citep{kruger15}.  In \cite{kruger15}, it is assumed that this particle had similar mechanical properties as those of a particular aerogel sample ($\rho=0.25\,g/cm^3$ and $Y=15\,MPa$) used for calibration purposes, given that, it was possible to match the detected $U_m$ and $T_c$ values with those  $U_m$ and $T_c$ values produced by the impact of aerogel particles in the laboratory (\citep{kruger15,  flandes15}. A summary of the properties of the detected particle is shown in Table~\ref{tab:PROP}. Other assumptions made for the calculation of the values of this table are that the detected grain was made out of water ice and that the actual surface of the comet may have a Young's modulus as large as $Y=500\,MPa$, which may set an upper limit for this property. 

%\begin{equation}  
%1.0<r<2.3\,{\rm mm}
%\label{EQ:4}
%\end{equation} 
%It was also derived that the particle had impacted the sensor with a speed in the interval:
%\begin{equation}  
%0.1< v<1.95\,{\rm m/s}
%\label{EQ:5}
%\end{equation} 

%Furthermore, assuming that the grain was made out of water ice its mass would be in the interval:
%\begin{equation}  
%10^{-6}< m<10^{-5}\,{\rm kg}
%\label{EQ:5}
%\end{equation} 

% its porosity was estimated as:	

%\begin{equation}  
%P=0.73
%\label{EQ:7}
%\end{equation} 
%or $73\%$.\\

%Which represents only a lower limit, since particles with larger Young's modulus would have a higher densities and smaller porosities. 

%-------------------------------------------------------------------------
%CHECK: ice values!

\begin{table}[tbp]
\large
   \centering
      \caption{Estimated properties of the particle detected by DIM }
   \begin{tabular}{@{} lcc @{}} % Column formatting, @{} suppresses leading/trailing space
   &&\\
      \hline
         \\   
      Density  [$g/cm^3$]         &$\rho$    &$0.25$   \\
      Radius  [$mm$]               &$r$         &$1.0-2.3$    \\
      impact speed  [$m/s$]     &$v$        &$0.1-1.95$    \\
      mass   [$kg$]                   &$m$      &$10^{-6}-10^{-5}$     \\
      Porosity                           &$P$&0.7\\
      \hline
   \end{tabular}
\label{tab:PROP}
\end{table}

%-------------------------------------------------------------------------

For comparison purposes, the Grain Impact Analyzer and Dust Accumulator (GIADA) onboard the spacecraft Rosetta detected few particles similar to DIM's except that at orbits of several to many tens of kilometers \citep{rotundi15}. This is further discussed in Section 2.

The goal in this paper is to present a simple, semi-analytic model for the gas flow from a comet which should help put the DIM and similar observations into a useful context. Detailed modeling of the gas and dust flow from the surface of a comet is a complex problem.  A proper calculation must take into account the highly non-spherical shape of the body and its rugged topography, as well as the fact that the surface is not uniformly covered with ice, not uniformly illuminated and has illumination varying with time.  In addition, there are considerable uncertainties in the interpretation of the observed signals \citep{flandes15}.  As a result, it can be helpful to apply a simple model that contains the essential physics in order to help understand and interpret the observations.  In what follows we investigate whether such a normal non-active surface is able to accelerate grains with properties similar to those
of the grain seen by DIM to their estimated speeds.

\section{Model}
We simplify the problem by assuming a smooth, spherical comet of radius $R$.  Let the gas production rate be $Q$ molec\,s$^{-1}$.  If the temperature of the gas is $T$, then the thermal velocity is given by
\begin{equation}
v_{th}=\sqrt{\frac{8kT}{\pi m}}
\end{equation}\label{vt}
where $k$ is Boltzmann's constant and $m$ is the mass of a gas molecule.  Although for the dilute gas at the comet's surface it is not strictly correct to assume a Boltzmann distribution, the approximation should be reasonable in this case.

One of the major difficulties is to determine the flow speed.  If a gas molecule undergoes many collisions near the surface, its direction of motion can be considered random, and the flow speed can be well-estimated by computing the average value of its radial motion.  This is typically $v_{th}/4$.  However, if the gas is very dilute and the molecules can travel a significant distance before their motion is randomized, most of the molecules will be moving away from the surface, and the flow speed can be higher.  In view of all the uncertainties involved, including the uncertainty in the temperature of the evaporating ice, $v_{th}/4$ seems to be a reasonable approximation at this point.

If $n$ is the number density of molecules, the flux leaving the surface can then be approximated by $${\cal F}=\frac{nv_{th}}{4}$$ so that 
\begin{equation}
Q=4\pi R^2{\cal F}=\pi R^2nv_{th}\label{Q}
\end{equation}

The drag force on a spherical grain of radius $a$ depends on the ratio of the molecular mean free path to the grain radius as given by the Knudsen number, which is defined by
$$Kn\equiv \frac{\lambda}{a}$$
Here $\lambda$ is the mean free path, given by
$$\lambda=\frac{1}{\sqrt{2}n\sigma}$$
where $\sigma$ is the molecular collision cross section.  For $Kn\gg 1$ Epstein drag \citep{epstein24} is appropriate, while for $Kn\ll 1$ one should use Stokes drag \citep{twomey77}.  A convenient way to interpolate between these two limits is via the Cunningham formula which gives the drag force on the grain as
\begin{equation}
F_{drag}=\frac{6\pi a \eta v}{\psi}\label{cunning}
\end{equation}
Here $\eta$ is the gas viscosity, $v$ is the relative velocity of the gas with respect to the grain, and $\psi$ is given by 
$$\psi=1+Kn\left[A+Be^{-C/Kn}\right]$$
Following \cite{allen82} we take $A=1.257$, $B=0.4$, and $C=1.1$.  It should be noted that while these values give a good fit to experiment, Eq.\,(\ref{cunning}) only reduces to the proper expression for Epstein drag when $A+B=1.5$, a condition which is not quite met here.

If the gas expands uniformly and flows with a constant velocity with respect to the comet nucleus, conservation of mass requires that the gas density at a distance $r$ from the surface be given by 
\begin{equation}
\rho_g(r)=\rho_g(R)\left(\frac{R}{r}\right)^2\label{gasden}
\end{equation}
The viscosity can be computed from kinetic theory \citep{loeb} to be
\begin{equation}
\eta=\frac{mv_{th}}{3\sqrt{2}\sigma}\label{visc}
\end{equation}
where $m$ is the mass of a gas molecule.  Equations \ref{cunning}, \ref{gasden}, and \ref{visc} can be used to compute the drag force on a grain at any point, provided the velocity of the grain is known.  This can be calculated from Newton's law by integrating
\begin{equation}
M_d\frac{du}{dt}=F_{drag}-F_{grav}\label{newton}
\end{equation}
where $u$ is the velocity of the grain with respect to the nucleus, $M_d$ is the mass of the grain, and $$F_{grav}=\frac{GM_dM_{comet}}{r^2}$$

These considerations can be put into a numerical context as follows: MIRO reports an average water production rate of $Q\sim 10^{26}$\,molec\,s$^{-1}$ (Ladislav Rezac, personal communication).  The mass of the grain is
$$M_d=\frac{4\pi}{3}\rho_da^3$$
The largest grain that can be lifted off the surface of the comet will be the grain for which 
$$F_{drag}=F_{grav}$$
at the surface.   For the very large mean free path at such low gas densities the drag force is given by the Epstein formula \citep{epstein24}
\begin{equation}
F_{drag}=\frac{4\pi}{3}a^2\rho_gvv_{th}\label{epstein}
\end{equation}
If we take $v=v_{th}/4$ and set $F_{drag}=F_{grav}$, it is easy to show that
 
\begin{equation}
a_{max}=\frac{3}{2\pi^2G}\frac{P_{vap}}{\rho_d\rho_c R}\label{amax1}
\end{equation}
where $P_{vap}$ is the vapor pressure of the sublimating gas, and $\rho_c$ is the density of the comet.  Alternatively one can write
\begin{equation}
a_{max}=\frac{Q}{2\pi^2G}\frac{\sqrt{2\pi mkT}}{\rho_dM_{comet}}\label{amax2}
\end{equation}
In the current semi-analytical model the grain must be slightly larger in order to be lifted off the surface because the Cunningham formula (Eq.\,(\ref{cunning})) does not exactly reproduce Eq.\,(\ref{epstein}) in the large Knudsen number limit.

\begin{figure}[btp]
\centerline{\includegraphics[width=20cm]{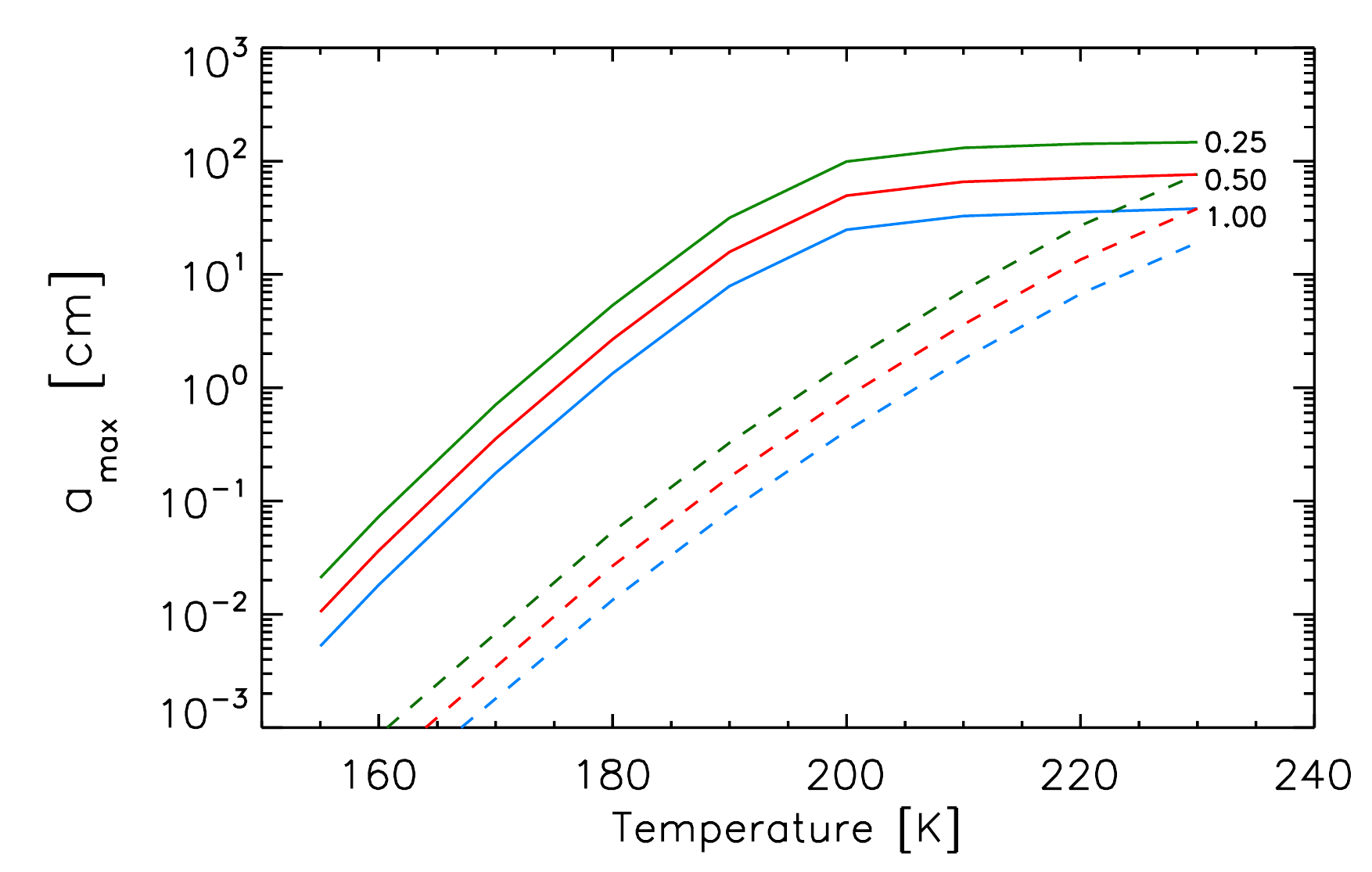}}
\caption{Maximum liftable radius as a function of surface temperature for three grain densities: 0.25 g\,cm$^{-3}$ (green), 0.5 g\,cm$^{-3}$ (red) and 1.0 g\,cm$^{-3}$ (blue).  Solid curves are for a pure ice surface, dashed curves are for the case where only 1\% of the surface is actively evaporating (see text). Note that grains larger than $\sim 5$\,cm will be larger than the DIM instrument and so are not directly relevant to the DIM measurment.}\label{fig:amax}
\end{figure}

For higher temperatures, the gas density can be high enough so that the Knudsen number falls below one.  In this case Eq.\,(\ref{amax2}) is no longer valid and $a_{max}$ must be found numerically.  The results are shown in Fig.\,\ref{fig:amax}.  The solid curves are for a surface of pure ice.  At low temperatures the gas density is low as well and the Knudsen number is large so that $\log a_{max}\propto T$ in accordance with Eq.\,(\ref{amax2}).  As the temperature increases, the Knudsen number decreases, and the slope of the curve changes so that $a_{max}$ becomes less sensitive to $T$.  If only a fraction, $f$, of the surface participates in evaporation, the total flux will decrease by the same factor and $Kn$ will remain large for higher temperatures.  In this case Eq.\,(\ref{amax2}) will hold for higher $T$.  This can be seen in the dashed curves in Fig.\,\ref{fig:amax}, which are for the case where only 1\% of the surface participates in evaporation.

If we take the density of the grain to be $\rho_d=0.25$\,g\,cm$^{-3}$ (see section 3, below), the largest grain that can be lifted off the surface with $Q=1.69\times 10^{26}$ (corresponding to $T=160$\,K) has a radius of 0.73\,mm.  Using a Monte Carlo code, \cite{tal14} calculated that the largest grain that could be lifted off from an emitting region under these conditions was between 0.18 and 0.25\,mm depending on how near the grain was to the edge of the emitting region of the comet.  However \cite{tal14} assumed a grain density of 1 g\,cm$^{-3}$.  From Eq.\,(\ref{amax2}) we see that the maximum radius of a liftable grain should be proportional to $\rho_d^{-1}$.  In this case the results of the Monte Carlo code should scale to 0.72 and 1.0\,mm, respectively.  This gives a measure of how reliable the current approximations are.  Indeed, for this case the mean free path is $\lambda=30$\,m which means that near the surface there are very few collisions and the distribution is far from Maxwell-Boltzmann.  The flow speed is probably higher than $v_{th}/4$ as we have assumed here, and the results from this simple model should be viewed with some caution.

\begin{figure}[btp]
\centerline{\includegraphics[width=20cm]{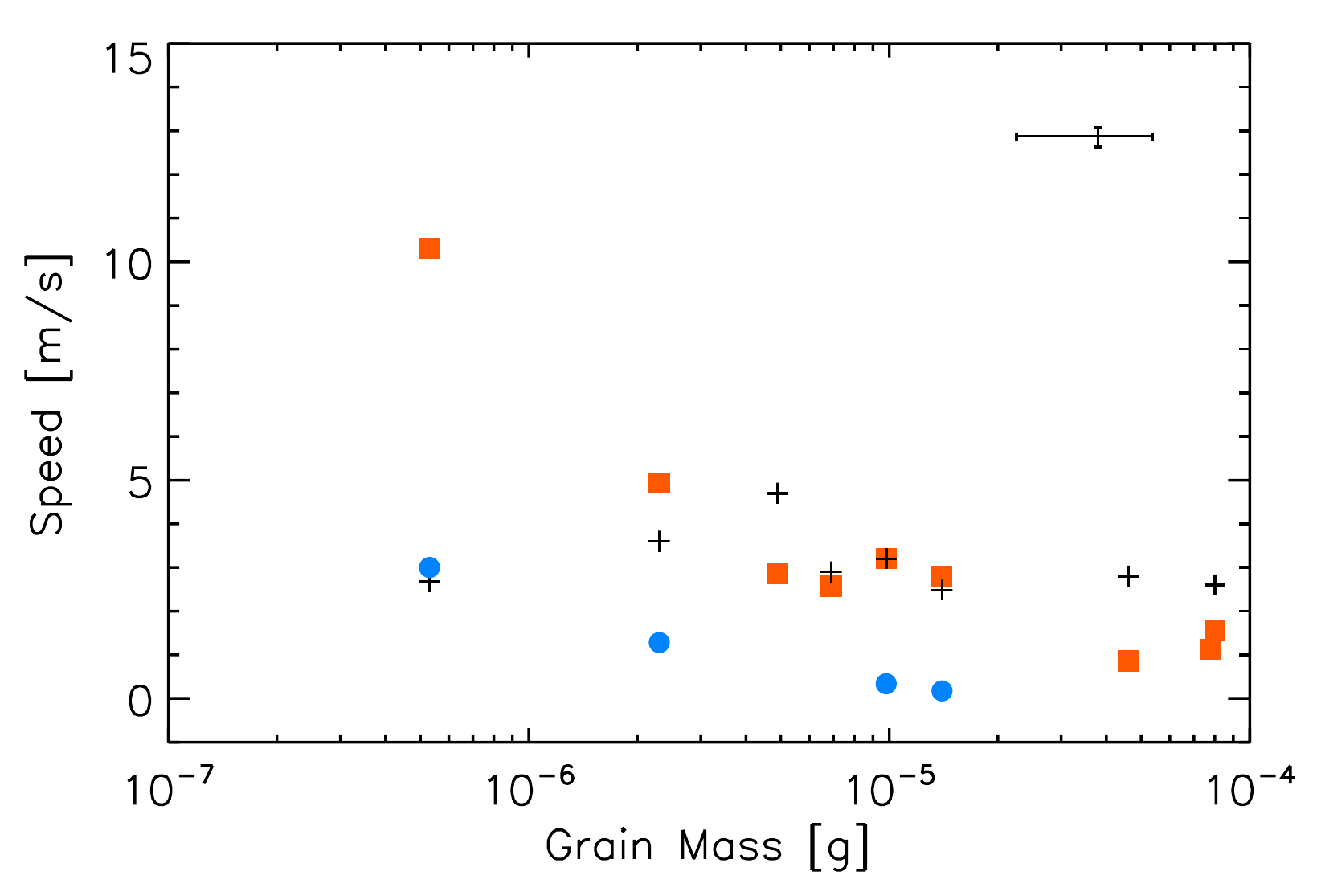}}
\caption{Speed as a function of mass for grains observed by GIADA at 30\,-\,92\,km distance from the nucleus.   Blue circles are for $Q=1.69\times 10^{26}$\,s$^{-1}$ ($T=160$\,K), red squares are for $Q=1.59\times 10^{27}$\,s$^{-1}$ ($T=170$\,K), black pluses are for the GIADA observations. Typical observational error bars for speed and mass are shown in the upper right. }\label{giada}
\end{figure}

As a second test of the model, we took the GIADA data given in Table 1 of \cite{rotundi15} with the exclusion of the low density event 33 (see footnote to Table 1 in \cite{rotundi15}).  These authors give the mass, velocity, and cross section of a number of grains that impacted at various distances (30\,-\,92\,km) from the comet.  Assuming a spherical grain, one can use the measured cross section to get the grain radius, and the mass to get the grain density.  Using these values in the current model we get the results  shown in Fig.\,\ref{giada}.

The blue circles represent the current model calculations assuming a production rate of $Q=1.69\times 10^{26}$\,s$^{-1}$ corresponding to a temperature of $T=160$\,K.  The red squares are for $Q=1.59\times 10^{27}$\,s$^{-1}$ corresponding to $T=170$\,K.  The black pluses are from Table 1 in \cite{rotundi15}.  Considering the uncertainties in the observations and in the model the results seem quite reasonable.

\begin{figure}
\centerline{\includegraphics[width=20cm]{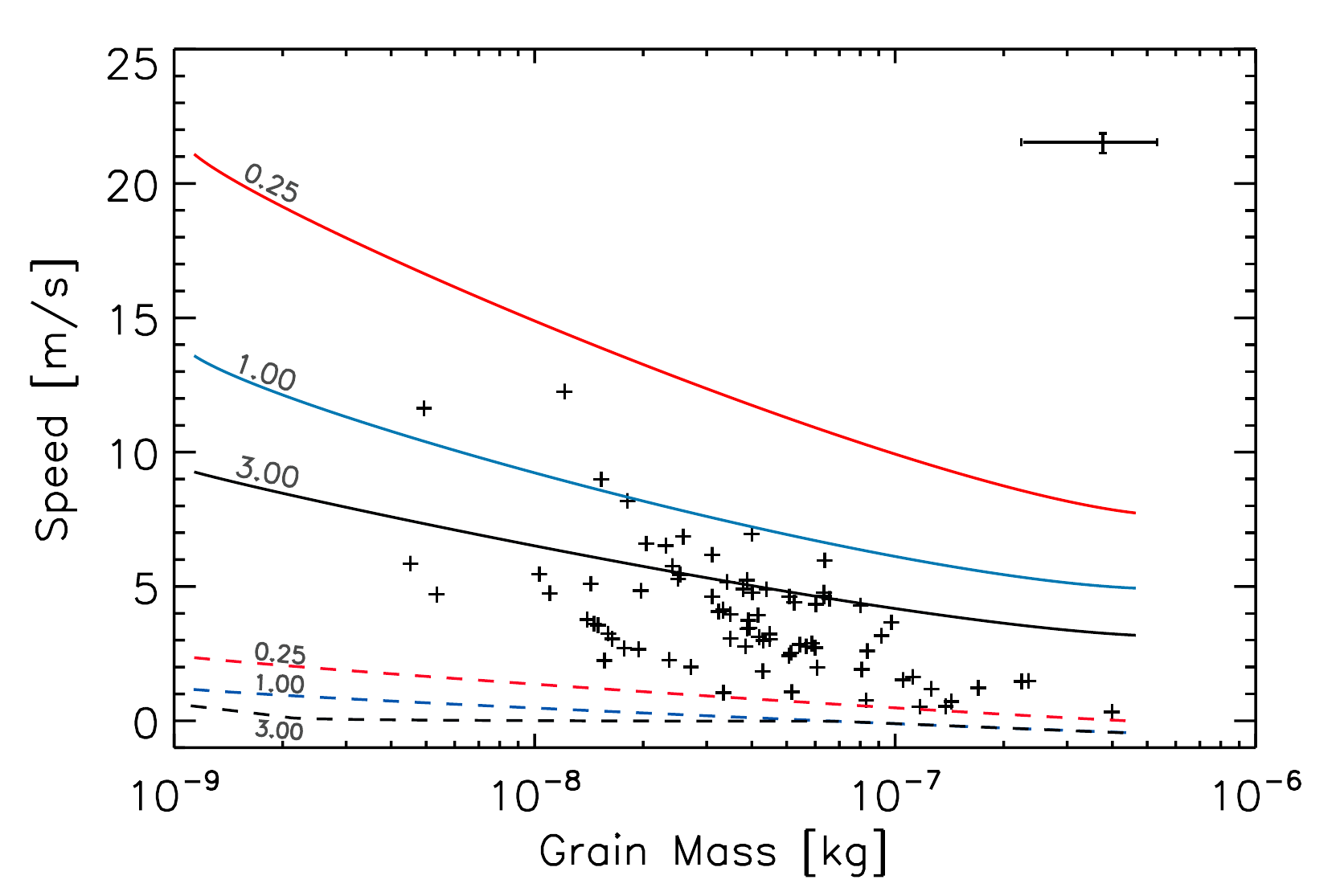}}
\caption{Grain speed as a function of grain mass at an altitude of 28\,km as measured by the GIADA instrument (black pluses).  Speeds predicted by the model are shown for three types of grain: rock (density = 3\,g\,cm$^3$-black curves), ice (density = 1\,g\,cm$^3$-blue curves) and porous ice (density = 0.25\,g\,cm$^3$-red curves) for assumed surface temperatures of 160\,K (dashed curves) and 180\,K (solid curves).  Typical observational error bars for speed and mass are shown in the upper right. }\label{fig:giada2}
\end{figure}
The model was evaluated for several surface temperatures and compared with the GIADA measurements of particle mass and velocity taken at distances between 10 and 30\,km from the center of the comet \citep{dellacorte2015}.  This additional data is shown in Fig.\,\ref{fig:giada2}. Here we have plotted the model predictions for the speed of grains at a distance of 28\,km from the surface as a function of their mass for surface temperatures of 160\,K (solid curves) and 180\,K (dashed curves).  The black pluses in the figure are the values observed by the GIADA instrument. The curves are for three assumed values of the grain density: 3\,g\,cm$^{-3}$, representing rocky material such as forsterite (black curves), 1\,g\,cm$^{-3}$, representing solid ice (blue curves), and 0.25\,g\,cm$^{-3}$ representing porous grains similar to that seen by the DIM instrument (red curves).  As can be seen from the figure, a surface temperature of 160\,K is too low to provide enough sublimation to reproduce the observed speeds even for very porous grains.  However a surface temperature of 180\,K can explain most of the observed grains even if they are composed of rocky material. 

In fact, almost all the GIADA measurements of mass and speed occurred in the northern hemisphere at latitude and longitude corresponding to the Neck (Hapi region) of the comet. In this region the surface temperatures reported by MIRO are higher than 180\,K \citep{schloerb2015}.  However, if only a fraction of the surface in this region has sublimating ice, only that fraction will contribute to the gas flux, and the total gas production rate will be correspondingly lower.  Thus a surface with 3.4\% ice coverage at 200\,K will have a water vapor production rate equivalent to a pure ice surface at 180\,K.  The flow speed of the water vapor will be somewhat different in the two cases, but the difference will be small enough to fall within the observational uncertainties in the measured grain speed and mass.  Ice coverage of a few percent of the surface is indeed consistent with the VIRTIS measurements in this region \citep{desanctis2015}.

Our model assumes that the sole driving force for accelerating dust grains away from the surface is the sublimation of water ice.  There are additional forces that should be considered.  Electrostatic forces have been shown to be effective in moving small grains \citep{wang09}, however these forces act mostly to levitate the grains some small distance from the surface. They are not capable of ejecting them from the comet.  The gas drag from a continuous flow into space is required to eject grains to distances of several kilometers from the nucleus.  

Our model also neglects any cohesive forces that may exist between grains.  Even for a rubble pile model of a comet \citep{blum14, blum15} such forces can generate tensile strengths of the order of 1 Pa \citep{gundlach15}.  In order to overcome these forces and eject grains, water ice must be heated to temperatures of the order of 215\,K.  Such temperatures are at the very upper limit of those observed for 67/P \citep{schloerb2015}.  Materials more volatile than water ice, such as CO$_2$, have been suggested as providing the force to overcome the cohesive forces \citep{gundlach15}, however the production rate of CO$_2$ (and other super-volatiles) is too low \citep{bockelee2015} for these species alone to accelerate the grains observed by DIM and GIADA to their estimated speeds.

\begin{figure}[btp]
\centerline{\includegraphics[width=20cm]{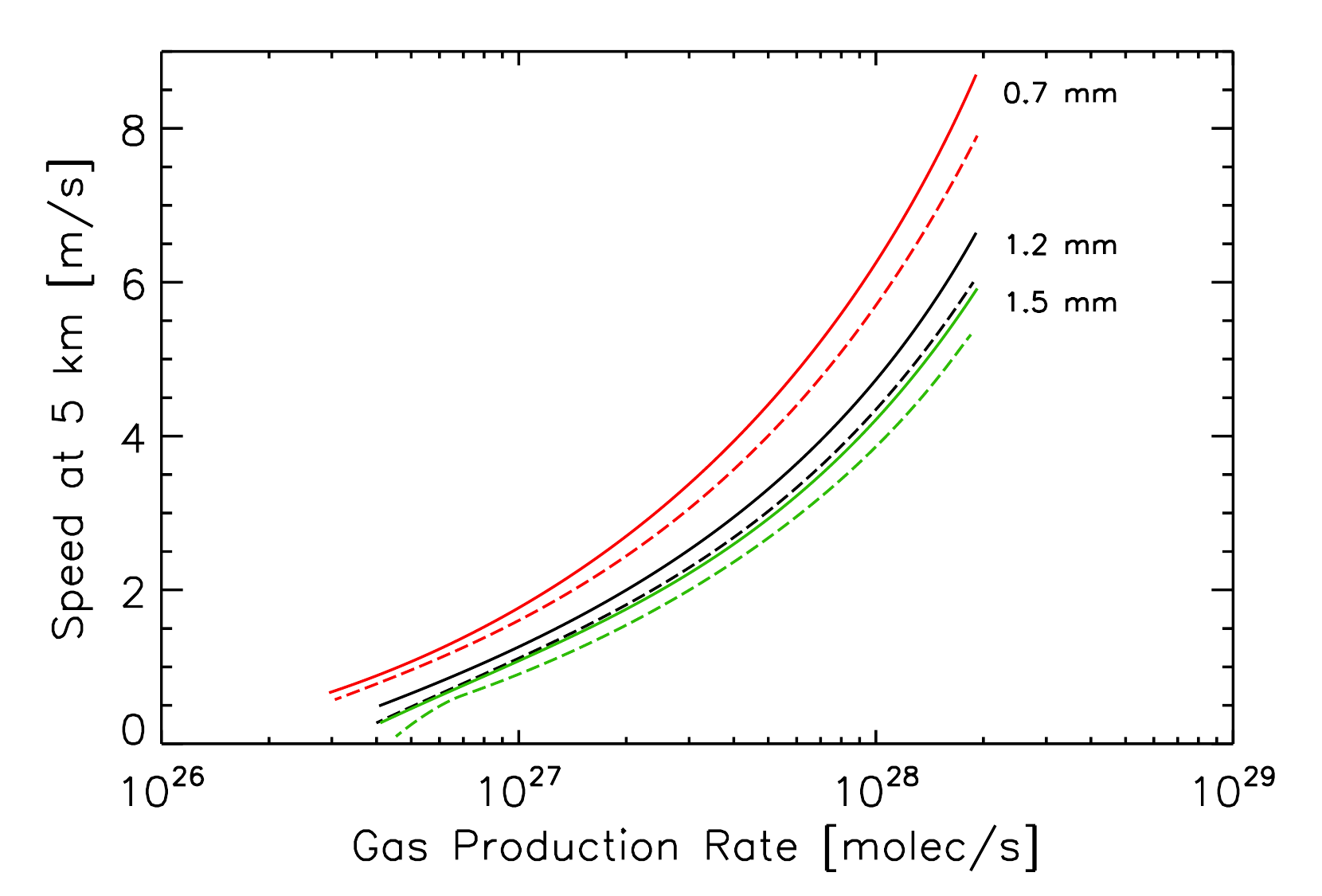}}
\caption{Speed at 5\,km from the center of the comet for a grain with $\rho_d=0.25$\,g\,cm$^{-3}$ as a function of $Q$.  Solid curves are for a density of 0.25\,g\,cm$^{-3}$ and dashed curves are for a density of 0.3\,g\,cm$^{-3}$.}\label{5kmspeed}
\end{figure}

\section{Conclusions}

From our semi-analytical model, Fig.\,\ref{5kmspeed} shows the speed at 5 km from the center of the comet as a function of $Q$ for three grain sizes assuming a density of $\rho_d = 0.25$\,g\,cm$^{-3}$ (solid curves). To check the sensitivity of our results, we also ran the same cases for $\rho_d = 0.3$\,g\,cm$^{-3}$ (dashed curves).  The difference is small, but noticeable.  We do not know the exact speed of the grain with respect to the comet, since we don't know the angle of impact with the DIM instrument. However this speed must be in the region of several m/s. As can be seen from the figure, if we assume that the force on the grain is due to the sublimation of water ice, then to achieve such a speed at a cometocentric distance of $5$\,km, the gas flow must correspond to a production rate of $Q\sim10^{27}$-$10^{28}$\,molec/s. This is one to two orders of magnitude higher than the average production rate reported by the MIRO instrument, and corresponds to ice sublimation temperatures of $\sim170-180$\,K. It seems likely that this grain was ejected by one of the more active regions where the local gas production rate was well above the average value.

The parameters of the particle detected by the DIM instrument can be understood in the context of a simple gas flow model.  This model is also consistent with the GIADA observations.  With a more detailed understanding of the fraction of the surface that is contributing to the evaporating gas and its temperature, we can further narrow down some of the free parameters of the model.  This, in turn, will help us to put better limits on the possible radius and density of the observed grains.

\section*{Acknowledgments}
This research was supported by the German
Bundesministerium fu\"r Bildung und Forschung through Deutsches Zentrum
fu ̈r Luft-und Raumfahrt e.V. (DLR, grant 50 QP 1302). A.F. is grateful
to MPS for financial support during a visit where part of this work was
done. A.F. was also supported by DGAPA through the Grant DGAPA-PAPIIT IA100114.
\newpage

\bibliographystyle{apalike}
\bibliography{mybibfile}

\end{document}